\documentstyle[12pt,cite,epsf]{article}
\let\Huge=\huge
\let\huge=\Large
\let\Large=\large
\let\large=\normalsize
\newcommand{\be}{\begin{equation}}
\newcommand{\ee}{\end{equation}}
\newcommand{\ba}{\begin{array}{c}}
\newcommand{\ea}{\end{array}}
\newcommand{\beqn}{\begin{eqnarray}}
\newcommand{\eeqn}{\end{eqnarray}}
\newcommand{\dis}{\displaystyle}
\newcommand{\bi}{\begin{itemize}}
\newcommand{\ei}{\end{itemize}}

\newcommand{\cO}{{\cal O}}

\newcommand{\lsim}{\stackrel{<}{_\sim}}
\newcommand{\gsim}{\stackrel{>}{_\sim}}
\newcommand{\toinf}{\stackrel{{}_{\large\sim}}{{}_{{}_{{}_{M_Q\to\infty}}}}}
\newcommand{\rms}{\rm\scriptsize}

\begin{document}
\begin{titlepage}
\vspace{0.2in}
\begin{flushright}
FTUV/94-37\\
IFIC/94-32\\
NBI-94-32\\
\end{flushright}
\vspace*{1.5cm}
\begin{center}
{\Huge \bf  Updating the Unitarity Triangle: \\
Top  Quark Mass Versus \\ Nonperturbative Uncertainties \\ }
\vspace*{0.8cm}
{\bf A. Pich$^{a)}$ and J. Prades$^{a,b)}$}\\
\vspace*{1cm}
$^{a)}$  Departament de F\'\i sica Te\`orica
and IFIC,  Universitat de Val\`encia -- CSIC\\
Dr. Moliner 50,
E--46100 Burjassot, Val\`encia, Spain
\\
$^{b)}$  Niels Bohr Institute and NORDITA, \\
Blegdamsvej 17, DK--2100 Copenhagen \O, Denmark \\
\vspace*{1.8cm}
{\Large\bf   Abstract  \\ }
\end{center}

We summarize the present knowledge on the
non-perturbative hadronic inputs needed in
the analysis of
$B^0$-$\bar B^0$ mixing and
the CP-violating parameter $\varepsilon$ of the $K^0$-$\bar K^0$
system.
Using this information, together with the recently determined value
of the top-quark mass,
we update the phenomenological constraints on the
unitarity triangle.
\vspace{4cm}
\begin{flushleft}
FTUV/94-37\\
IFIC/94-32\\
NBI-94-32\\
July 1994
\end{flushleft}
\end{titlepage}


\section{Introduction}

In the Standard Model \cite{ref:SM}, the GIM suppression
of flavour-changing neutral-current processes
\cite{ref:GIM}
and the necessarily presence of three quark
families to generate CP-violation effects make
the top quark a key ingredient to analyze these phenomena.
In both cases, the unitarity of the
Cabibbo--Kobayashi--Maskawa (CKM) matrix implies vanishing effects
in the limit of degenerate quark masses.
Those processes, occurring through one-loop diagrams,
are then very sensitive to the masses of the three
equal-charge quarks running along the internal lines.
Due to its large mass, the top-quark gives a very important
(often dominant) contribution; thus, the unknown value of $m_t$ has
been up to now a crucial uncertainty in the phenomenological analyses.

The recent announcement of evidence for the top quark \cite{ref:CDF94},
with a ``pole'' mass
\begin{equation}
\label{eq:mt_cdf}
   m_t \, = \, 174 \pm 10 {}^{+13}_{-12} \,\mbox{\rm GeV \qquad\qquad
        (CDF),}
\end{equation}
should allow to improve the present determinations of the CKM parameters.
This value of the top mass is in excellent agreement with the
range obtained from the Standard Model electroweak fits at the
$Z$ peak \cite{ref:LEP94},
\begin{equation}
\label{eq:mt_lep}
        m_t \, = \, 177 {}^{+11}_{-11} {}^{+18}_{-19}
\,\mbox{\rm GeV \qquad\qquad (Electroweak Fits),}
\end{equation}
which gives further support to the CDF analysis.

\begin{figure}
\vspace*{-4cm}
{\epsfysize=17cm\epsfxsize=14.5cm\epsfbox{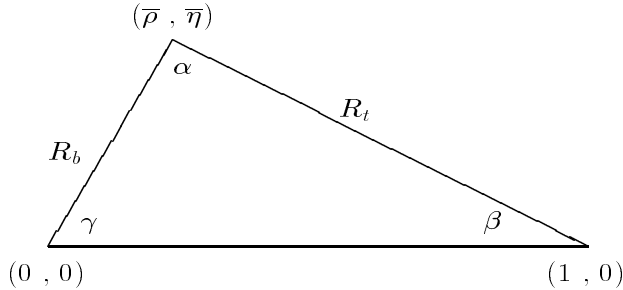}}
\vspace*{-8cm}
\caption{Unitarity triangle}
\label{fig:triangle}
\end{figure}

One of the crucial tests of the Standard Model mechanism of CP
violation involves the unitarity condition
\begin{equation}
\label{eq:unitarity}
        V_{ub}^* V_{ud}^{\phantom{*}} + V_{cb}^* V_{cd}^{\phantom{*}} +
        V_{tb}^* V_{td}^{\phantom{*}} \, = \, 0 .
\end{equation}
This relation can be visualized as a triangle in the complex plane,
the so-called ``unitarity triangle'' (UT).
In the absence of CP violation, the triangle would degenerate
into a segment along the real axis.
It has become conventional to scale the triangle, dividing its sides
by $|V_{cb}^* V_{cd}^{\phantom{*}}|$.
In the Wolfenstein parametrization \cite{ref:WO83}
of the CKM matrix,
\be\label{eq:wolfenstein}
{\bf V}\, =\,
\left[ \matrix{\displaystyle \ 1- {\lambda^2 \over 2}
\hfill&
\displaystyle \ \ \ \ \ \ \lambda \hfill&
\displaystyle \ \ \ \ \ A\lambda^
3(\rho  - i\eta) \hfill \cr\displaystyle
\hfill& \displaystyle \hfill&
\displaystyle \hfill \cr\displaystyle \ \ \ -\lambda
\hfill& \displaystyle \ \
\ \ \ 1 -{\lambda^ 2 \over 2} \hfill& \displaystyle
\ \ \ \ \ A\lambda^ 2 \hfill
\cr\displaystyle \hfill& \displaystyle \hfill&
\displaystyle \hfill
\cr\displaystyle \ A\lambda^ 3(1-\rho -i\eta)
\hfill& \displaystyle \ \ \ \ \
-A\lambda^ 2 \hfill& \displaystyle
\ \ \ \ \ \ \ \ \ 1 \hfill \cr} \right]\
+\ O\left(\lambda^ 4 \right) \, ,
\ee
$|V_{cb}^* V_{cd}^{\phantom{*}}|$ is real to an excellent accuracy
[$O(\lambda^7)$]. Therefore, the scaling aligns one side of the triangle
along the real axis and makes its length equal to 1;
the coordinates of the 3 vertices are then (0,0), (1,0) and
$(\bar\rho,\bar\eta)$, with \cite{ref:BLO94}
\begin{equation}
\label{eq:bar_eta_rho}
        \bar\rho\, \equiv \, \rho\left(1-{\lambda^2\over 2}\right) ,
        \qquad
        \bar\eta\, \equiv \, \eta\left(1-{\lambda^2\over 2}\right) .
\end{equation}
The triangle is shown in Fig.~\ref{fig:triangle}.

The sides of this triangle,
\beqn
\label{eq:rb}
R_b\,\equiv\,
\left|{V_{ub}^* V_{ud}^{\phantom{*}} \over V_{cb}^*
V_{cd}^{\phantom{*}}}\right|
\, = \, \left(1-{\lambda^2\over 2}\right)
\left| {V_{ub}\over\lambda V_{cb}}\right| \, = \,
\sqrt{\bar\rho^2 + \bar\eta^2} ,
\\ \label{eq:rt}
R_t\,\equiv\,
\left|{V_{tb}^* V_{td}^{\phantom{*}} \over V_{cb}^*
V_{cd}^{\phantom{*}}}\right|
\, = \,
\left| {V_{td}\over\lambda V_{cb}}\right| \, = \,
\sqrt{(1-\bar\rho)^2 + \bar\eta^2} , \qquad\;\,
\eeqn
can be determined through CP-conserving measurements:
the ratio $\Gamma(b\to u)/\Gamma(b\to c)$ fixes $R_b$, while
$R_t$ can be extracted from the observed $B^0_d$-$\bar B^0_d$ mixing.
A third constraint is obtained from the measured value of the
CP-violation parameter $\varepsilon$.

A precise test of the unitarity relation (\ref{eq:unitarity}) is
obviously required. Moreover,
the angles $\alpha$, $\beta$ and $\gamma$ of the UT regulate many
interesting CP-violating asymmetries in the $B$ system
(see for instance \cite{ALE93}); thus, it
is important to determine the triangle, in order to know the
expected size of the CP-signals at future $B$ factories.

The theoretical analysis of the UT is quite straightforward and
has been performed many times in the past. Nevertheless,
since the constraints obtained from $\varepsilon$ and from
$B^0_d$-$\bar B^0_d$ mixing strongly depend on the value of the
top mass, the determination of the UT needs to be updated in view
of the new information provided by CDF.
A first study \cite{ref:BLO94}, ``guessing'' the CDF value before
its public announcement, has already been done, and
a second one \cite{ref:AL94} has just appeared, immediately after
the CDF publication.
Many more are probably going to show up soon.

Unfortunately, the knowledge of the top-quark mass is not enough to
precisely fix the UT.
Our limited ability to handle the long-distance effects of the strong
interactions,
translates into unavoidable hadronic uncertainties, which enter in the
determination of the CKM parameters.
Although a big theoretical effort has been made during the last decade
to study the relevant hadronic matrix elements, the present status is
certainly not satisfactory: theoretical errors are still large, and
there is no universal agreement on the values of some
non-perturbative parameters.

The resulting uncertainties are not always properly reflected in the
usual UT analyses. Quite often, the non-perturbative parameters are fixed
in a rather ad-hoc way, or following the last fashion (i.e. taking
similar values to the latest published analyses);
thus, the mere repetition
of a parameter set, rather than the scientific
quality, is what finally makes a given choice of non-perturbative inputs
the one ``considered more reliable''.
A very instructive example is provided by the
$B^0$-$\bar B^0$ mixing parameter $\xi_B\equiv f_B\sqrt{B_B}$:
while in 1988 it had been already established\footnote{
A summary of existing calculations of $f_B$, $B_B$ and $\xi_B$
was given in
Ref.~\protect\cite{ref:PI90}.}
that $\xi_B > f_\pi$ \cite{ref:PI88},
many UT analyses done from 1988 to 1991
were still using $\xi_B < f_\pi$ and getting, therefore,
meaningless results.

In the following, we want to critically summarize the present status
of those hadronic matrix elements which are relevant for the
UT determination, and work out their phenomenological implications.
We will try to put forward the
arguments supporting our final choice of parameters and their
associated error-bars.
Using the measured value of the top-quark mass \cite{ref:CDF94,ref:LEP94},
together with the most recent experimental information on the $B$ system
\cite{ref:DA93,ref:VE94,ref:cleo93,ref:ST94,ref:cleo94},
we will finally analyze the present constraints on the UT.

\section{Master formulae}

The short-distance analysis of $K^0$-$\bar K^0$ and $B^0$-$\bar B^0$ mixing
is well known. An excellent review has been given by Buras and Harlander
\cite{ref:BH92}. Here, we only list the final formulae relevant for our
discussion, referring to Refs.~\cite{ref:BLO94,ref:BH92} for further
details.

The experimental value of $\varepsilon$ specifies a hyperbola in the
$(\bar\rho,\bar\eta)$ plane:
\begin{equation}
\label{eq:hyperbola}
        \bar\eta\left[(1-\bar\rho) A^2\widehat{\eta}_2
S(r_t) + P_0\right] A^2 \widehat{B}_K
        \, = \, {3\sqrt{2}\pi^2 \Delta M_K |\varepsilon| \over
        G_F^2 M_W^2 f_K^2 M_K \lambda^{10}} \, \equiv \, C_\varepsilon ,
\end{equation}
where
\be
\label{eq:st}
S(r_t) \, = \, {r_t \over 4} \left[ 1 + {9 \over 1 - r_t}
- {6 \over (1-r_t)^2} - {6 r_t^2 \ln{r_t}\over (1-r_t)^3}\right]
\ee
contains the dominant top contribution, and the corrections
coming from the $cc$ and $tc$ box diagrams are given by
\beqn\label{eq:p0}
P_0 &=& {1\over \lambda^4}\left[\widehat{\eta}_3 S(r_c,r_t)-
\widehat{\eta}_1r_c\right],
\\ \label{eq:sct}
S(r_c,r_t) &=&  r_c \left[\ln{\left(r_t\over r_c\right)}
- {3 r_t \over 4 (1 - r_t)}
\left( 1 + {r_t \ln{r_t}\over 1 - r_t}\right)\right] .
\eeqn
Here, $r_q\equiv m_q^2/M_W^2$
and the renormalization-scale-invariant factors
\be\label{eq:eta_factors}
\widehat{\eta}_1 = 1.10, \qquad
\widehat{\eta}_2 = 0.57, \qquad
\widehat{\eta}_3 = 0.36,
\ee
take into account the computed short-distance QCD corrections.

The main theoretical uncertainty stems from the so-called
$\widehat{B}_K$ factor,
parametrizing the hadronic matrix element of the $\Delta S=2$
four-quark operator:
\beqn\label{eq:bk}
\langle\bar K^0|\left(\bar s\gamma^\mu(1-\gamma_5)d\right)
\left(\bar s\gamma_\mu(1-\gamma_5)d\right)|K^0\rangle
\,\equiv\, {8\over 3} \left( \sqrt{2} f_K M_K\right)^2
B_K(\mu^2) ,
\\ \label{eq:bkhat}
\widehat{B}_K \,\equiv\, \alpha_s(\mu^2)^{-2/9}\, B_K(\mu^2) .
\qquad\qquad\qquad\qquad
\eeqn
The hadronic matrix element (and thus $B_K$) depends on the
chosen renormalization scale; this dependence is exactly cancelled
by the short-distance renormalization-group factor
$\alpha_s(\mu^2)^{-2/9}$, so that the combination $\widehat{B}_K$
appearing in (\ref{eq:hyperbola}) is renormalization-scale
independent.

In the neutral $B$ meson system, the mixing is completely dominated by the
top contribution:
\be\label{eq:b_mixing}
x_d \,\equiv\, {\Delta M_{B_d}\over\Gamma_{B_d}} \, = \,
\tau_{B_d} |V_{td}|^2 M_{B_d} {G_F^2 M_W^2\over 6\pi^2}
\widehat{\eta}_B S(r_t) (\sqrt{2} f_B)^2 \widehat{B}_B .
\ee
The short-distance QCD-correction is collected in the
the renormalization-scale-invariant factor $\widehat{\eta}_B$,
which has been computed to be
\be\label{eq:eta_b}
\widehat{\eta}_B = 0.55 ,
\ee
and the long-distance $\Delta B=2$
hadronic matrix element
is parametrized in terms of
\be\label{eq:xi_b}
\hat\xi_B \,\equiv\, f_B \sqrt{\widehat{B}_B}
\,\equiv\,\alpha_s(\mu^2)^{-3/23}\,\xi_B(\mu^2) ,
\ee
where the renormalization-scale-independent factor
$\widehat{B}_B$ is
defined by
\beqn\label{eq:bb}
\langle\bar B^0|\left(\bar b\gamma^\mu(1-\gamma_5)d\right)
\left(\bar b\gamma_\mu(1-\gamma_5)d\right)|B^0\rangle
\,\equiv\, {8\over 3} \left( \sqrt{2} f_B M_B\right)^2
B_B(\mu^2) ,
\\ \label{eq:bbhat}
\widehat{B}_B \,\equiv\, \alpha_s(\mu^2)^{-6/23}\, B_B(\mu^2) ,
\qquad\qquad\qquad\qquad
\eeqn
in complete analogy
to Eqs.~(\ref{eq:bk}) and (\ref{eq:bkhat}).
The different power of $\alpha_s(\mu^2)$ in
Eqs. (\ref{eq:bkhat})  and (\ref{eq:bbhat}) is due
to the different number of ``light'' quark flavours (3 and 5,
respectively) in the $K$ and $B$ systems.

The QCD parameters $\widehat{\eta}_1$, $\widehat{\eta}_2$ and
$\widehat{\eta}_B$ have been computed at the next-to-leading-logarithm
order \cite{ref:BLO94}.
At this level of accuracy, one needs to state how $m_t$ is defined.
The numerical values quoted in Eqs. (\ref{eq:eta_factors}) and
(\ref{eq:eta_b})
correspond to the running top quark mass in the $\overline{MS}$ scheme
evaluated at $m_t$, i.e. in Eqs. (\ref{eq:hyperbola}) to (\ref{eq:bbhat})
$m_t$ stands for $\overline{m}_t(m_t)$ \cite{ref:BLO94}.
The relation with the ``pole'' mass, defined as the pole of the
renormalized propagator,
 is given by \cite{ref:NBGS}
\beqn \label{eq:pole_mass}
m_t^{\mbox{\rms pole}} & = &
\overline{m}_t(m_t^{\mbox{\rms pole}}) \,
\left\{ 1 + {4 \over 3} {\alpha_s(m_t^{\mbox{\rms pole}})
\over \pi}\vphantom{\left({\alpha_s(m_t^{\mbox{\rms pole}})
\over \pi} \right)^2 } \right.
\nonumber \\  & + & \left. \left[ 16.11 -1.04 {\dis \sum_{i=u,d,s,c,b}}
\left(1- {m_i^{\mbox{\rms pole}} \over m_t^{\mbox{\rms pole}}}
\right)\right] \left({\alpha_s(m_t^{\mbox{\rms pole}})
\over \pi} \right)^2 \right\} .
\eeqn
Thus, $\overline{m}_t(m_t)$ is about 9 GeV lower than
$m_t^{\mbox{\rms pole}}$.
The measured values (\ref{eq:mt_cdf}) and (\ref{eq:mt_lep})
should be identified with $m_t^{\mbox{\rms pole}}$.

\section{Experimental inputs}

The physical quantities defining the parameter $C_\varepsilon$
are rather well measured \cite{ref:PDG92}:
\be\label{eq:k_inputs}
\begin{array}{ll}
G_F = 1.16639 (2) \times 10^{-5} \,\mbox{\rm GeV}^{-2} ,
&
M_W = 80.22 \pm 0.16 \,\mbox{\rm GeV} , \qquad
\\
\Delta M_K = (3.522\pm0.016)\times 10^{-12}\,\mbox{\rm MeV},
&
M_{K^0} = 497.671\pm 0.031 \,\mbox{\rm MeV},
\\
f_K = 1.22\, f_\pi = 113\pm 1 \,\mbox{\rm MeV},
&
\lambda = 0.2205\pm 0.0018 ,
\\
|\varepsilon| = (2.26\pm 0.02)\times 10^{-3} .
\end{array}
\ee
Therefore,
\be\label{eq:c_epsilon}
C_\varepsilon \, = \, 0.220\pm 0.019 .
\ee
Owing to its large power ($\lambda^{-10}$),
the dominant uncertainty comes from the value of $\lambda$.

Except for the $B^0$ meson mass
\cite{ref:PDG92},
\be\label{eq:mB0}
M_{B^0}\, = \, 5.279 \pm 0.002 \, \mbox{\rm GeV} ,
\ee
the experimental error-bars
are somewhat larger in the $B$ system.
The averaged value of the $b$-lifetime has been continuously
increasing as function of time; the 1990 value \cite{ref:PDG90}
$\langle \tau_b\rangle = 1.18\pm 0.11 \,\mbox{\rm ps}$
moved up to
$\langle \tau_b\rangle = 1.29\pm 0.05 \,\mbox{\rm ps}$
in 1992 \cite{ref:PDG92}, and has been further increased by the
recent LEP data.
The present world average \cite{ref:DA93,ref:VE94} is:
\be\label{eq:b_lifetime}
\langle \tau_b\rangle = 1.49\pm 0.04 \,\mbox{\rm ps} .
\ee
This value is in good agreement with the measured ``exclusive''
lifetime of the $B^0$ meson \cite{ref:DA93,ref:VE94},
\be\label{eq:BM_lifetime}
\langle\tau(B^0)\rangle = 1.5\pm0.1 \,\mbox{\rm ps} .
\ee

The world averaged value of the $B^0$-$\bar B^0$ mixing parameter
is \cite{ref:DA93}:
\be\label{eq:xd}
x_d = 0.71\pm 0.07 .
\ee

The experimental determination of the CKM matrix elements
$|V_{cb}|$ and $|V_{ub}|$ requires theoretical input and,
therefore, suffers from systematic uncertainties related to
the model-dependence involved in the analysis of semileptonic
$b$ ($B$) decay. The cleanest determination of $|V_{cb}|$
uses the decay $B\to D^* l \bar\nu_l$ \cite{ref:NE91}, where
the relevant hadronic form factor ($h_{A_1}$)
can be controlled at the
level of a few per cent, close to the zero-recoil region.
In the infinite $B$-mass limit, the normalization of this
form factor at zero recoil is fixed to be one, and the leading
$1/M$ corrections  vanish \cite{ref:LU90} due to
heavy-quark symmetry; thus, the theoretical uncertainty is
of order $1/M^2$ and therefore in principle small. The
calculated short-distance QCD corrections \cite{ref:NE93}
and the present estimates
of the $1/M^2$ contributions \cite{ref:FN94}  
result in $h_{A_1}(1) = 0.97\pm 0.04$ \cite{ref:NE94}, 
implying \cite{ref:cleo94}
\be\label{eq:V_cb}
|V_{cb}| \left({\tau(B_d^0)\over 1.5 \,\mbox{\rm ps}}\right)^{1/2}
\, = \,
0.037 \pm 0.004.
\ee

An independent determination can be done
using the inclusive semi-leptonic decay spectra. The theoretical
uncertainty is however bigger, because the total rate
scales as $m_b^5$ and, therefore, is very sensitive to the
not so-well-known value of the b-quark mass.
A compilation of experimental analyses \cite{ref:ST94} finds
$|V_{cb}| \left(\tau(B_d^0)/ 1.5 \,\mbox{\rm ps}\right)^{1/2}
=
0.040 \pm 0.005$, in good agreement with (\ref{eq:V_cb});
however, the quoted error does not take into account the
uncertainty associated with $m_b$.
A recent study of the $b$-quark mass definition within
the heavy quark effective theory finds a slightly
heavier $b$ quark and consequently a slightly smaller value
$|V_{cb}| \left(\tau(B_d^0)/ 1.5 \,\mbox{\rm ps}\right)^{1/2}
=
0.036 \pm 0.005$ \cite{ref:BN94}.

The value of $|V_{cb}|$ quoted in Eq. (\ref{eq:V_cb}) implies,
\be\label{eq:a}
A \, = \, 0.76\pm 0.08 .
\ee

The present determination of $|V_{ub}|$  is based on
measurements of the lepton momentum spectrum in inclusive
$B\to X_ql\bar\nu_l$ decays, where $X_q$ is any hadronic state
containing a quark $q=c$ or $u$.
The method is very sensitive to the assumed theoretical spectrum
near the kinematic limit for $B\to D l\bar\nu_l$.
Using different models to estimate the systematic theoretical
uncertainties, the analyses of the experimental data give
\cite{ref:DA93,ref:cleo93}:
\be
\left|{V_{ub}\over V_{cb}}\right| \, = \, 0.08\pm 0.02.
\ee

\section{$\widehat{B}_K$ factor}

The matrix element (\ref{eq:bk}) was first estimated \cite{ref:GL74}
via the assumption of vacuum saturation, i.e.
splitting the matrix element in a product of two
currents and
inserting the vacuum in all possible ways.
The factor $B_K$ parametrizes the deviation from this
factorization estimate, so that
\be\label{eq:bk_vs}
B_K(\mu^2)\, = \, 1 \qquad\qquad \mbox{\rm (Vacuum Saturation)}
\ee
corresponds to the vacuum saturation approximation.
Clearly, this approximation can only be taken as an
order-of-magnitude estimate, since it completely ignores
the renormalization group factor $\alpha_s(\mu^2)^{-2/9}$.
As it stands, Eq.~(\ref{eq:bk_vs}) is meaningless
because the value of $\mu$ is not specified.

An improved factorization estimate of
the renormalization-scale-independent
factor $\widehat{B}_K$ can be trivially performed
in the large-$N_c$ limit, where $N_c$ is the number of colours.
In this limit, the anomalous dimension of the $\Delta S=2$ operator
vanishes and factorization is then exact \cite{ref:GTW81}:  
\be\label{eq:bk_largeN}
\widehat{B}_K \, = \, B_K \, = \, {3\over 4}
\qquad\qquad (N_c\to\infty) .
\ee

Another approach allowing a rigorous calculation of $\widehat{B}_K$,
within a well-defined approximation, is Chiral Perturbation
Theory (CHPT).
The $\Delta S=2$ operator has the same chiral transformation
properties than the four-quark operator mediating $\Delta S=1$,
$\Delta I=3/2$ transitions. Both operators belong to the same
$(27_L,1_R)$ representation of the chiral
$SU(3)_L\otimes SU(3)_R$ group and, therefore,
their matrix elements are
trivially related by a Clebsch-Gordan coefficient.
One can then directly measure the value of $\widehat{B}_K$,
using the $K^+\to\pi^+\pi^0$ decay rate \cite{ref:DGH82}.
One gets in this way
\be\label{eq:bk_p2}
\widehat{B}_K\, = \, 0.37 \qquad\qquad (\mbox{\rm CHPT at } O(p^2)).
\ee
At lowest-order in Chiral Perturbation Theory, $O(p^2)$,
the only possible corrections to this result are $SU(2)$ violations,
induced by electromagnetism or proportional to $m_d-m_u$
\cite{ref:RA89}.
$SU(3)$-breaking effects, spoiling the relation
with the $K^+\to\pi^+\pi^0$ rate, appear first at the next order
in the momentum expansion, i.e. at $O(p^4)$.

The discrepancy between the two determinations of $\widehat{B}_K$
in Eqs.~(\ref{eq:bk_largeN}) and (\ref{eq:bk_p2})
shows that higher-order corrections should be sizeable.
A first estimate of the $1/N_c$ corrections to (\ref{eq:bk_largeN})
was performed by Bardeen, Buras and G\'erard \cite{ref:BBG88},
with the result:
\be\label{eq:bk_bbg}
\widehat{B}_K\, = \, 0.70\pm 0.10 \qquad\qquad
(\mbox{\rm Next-to-Leading }  1/N_c\,
\mbox{\rm  Estimate}).
\ee
The fact that the next-to-leading corrections in the $1/N_c$
expansion are negative was further demonstrated
in Ref. \cite{ref:PR91}, using functional integration techniques.
The numerical result of this analysis
\be\label{eq:bk_pr}
\widehat{B}_K\, = \, 0.4\pm 0.2 \qquad\qquad
(\mbox{\rm Next-to-Leading }  1/N_c\,
\mbox{\rm  Estimate, } O(p^2)),
\ee
has a rather large error, but the negative sign of the $1/N_c$
corrections was clearly established  \cite{ref:PR91}.
At $O(p^2)$, this sign can in fact be proven to be negative
in a model-independent way \cite{ref:PR94}, because the
$1/N_c$ correction to $\widehat{B}_K$ is anticorrelated with
the one enhancing the $\Delta I=1/2$ $K\to\pi\pi$ amplitude.

Since $f_K/f_\pi=1.22$, $SU(3)$ breaking
corrections to the result (\ref{eq:bk_p2}) could be expected
to be important.
The $SU(3)$ rotation of the $K^+\to\pi^+\pi^0$ amplitude determines
the product $\hat\xi_K^2\equiv f_K^2 \widehat{B}_K$.
The value in Eq.~(\ref{eq:bk_p2}) is obtained fixing (by definition)
the kaon decay constant to its physical value; however,
at $O(p^2)$ in the chiral expansion,
$f_K = f_\pi$. Using instead the pion decay constant,
the same value of $\hat\xi_K$ results in a 40\% bigger $\widehat{B}_K$
factor:
$\widehat{B}_K = 0.37 (f_K/f_\pi)^2 = 0.55$.
In fact, those higher-order chiral corrections which are
factorizable
will precisely change $f_\pi$ to $f_K$,
increasing the lowest-order result (\ref{eq:bk_p2}) by the
factor $(f_K/f_\pi)^2$ and making it closer to
the leading $1/N_c$ estimate (\ref{eq:bk_largeN}).
Of course, the relevant question now concerns the magnitude
and sign of the non-factorizable $O(p^4)$ chiral contributions.

The explicit calculation of the 1-loop chiral
corrections shows \cite{ref:BSW84} indeed that
the $\Delta S=2$ $K^0$-$\bar K^0$ matrix element
(i.e. $\hat\xi_K$) receives a large (and positive) logarithmic
correction; a big part of this 1-loop contribution is just
the usual (factorizable) correction to $f_K^2$.
To perform a complete $O(p^4)$ CHPT calculation, one needs also
the non-logarithmic contributions coming from next-to-leading
terms in the chiral weak Lagrangian.
A recent estimate \cite{ref:BR94},  based on the
$1/N_c$ techniques of Ref.~\cite{ref:PR91},
finds that the non-factorizable chiral corrections are negative.
Adding all contributions, the final result of Ref. \cite{ref:BR94}
is:
\be\label{eq:bk_p4}
\widehat{B}_K \, = \, 0.42\pm0.06
\qquad\qquad
(O(p^4)\,\mbox{\rm CHPT } + 1/N_c\,\mbox{\rm  Estimate}).
\ee

Two clear qualitative conclusions emerge from the previous
analyses:
\begin{itemize}
\item The next-to-leading corrections in the $1/N_c$ expansion
(non-factorizable corrections)
are negative and, therefore, decrease
the leading-order result (\ref{eq:bk_largeN}).
\item The factorizable $O(p^4)$ corrections increase
the lowest-order CHPT result
(\ref{eq:bk_p2}); however, the non-factorizable contributions
appear to be negative.
\end{itemize}
Thus, quite independently of any particular numerical estimate,
one can pin down $\widehat{B}_K$ to be within the interval:
\be\label{eq:bk_general}
0.35 \, <  \,\widehat{B}_K \, < \, 0.75 .
\ee

$\widehat{B}_K$ can be also calculated, through
dispersion relations, using a QCD-hadronic duality approach.
Making a dual description of the $\Delta S=2$ operator
(CHPT + hadronic resonances at long distances and the usual
quark-gluon description at short distances) and analyzing
the corresponding
two-point function correlator, it is possible to extract the
value of the $K^0$-$\bar K^0$ matrix element \cite{ref:PR85}.
Updating all the inputs of this analysis, one gets
\cite{ref:PDPPR91,ref:JM94}
\be\label{eq:bk_2pf}
\widehat{B}_K \, = \, 0.39\pm0.10
\qquad\qquad\mbox{\rm (QCD-Hadronic Duality)}.
\ee
This result is in fact a calculation of the relevant
$O(p^2)$ chiral coupling; thus, one could probably expect a
small increase, due to $O(p^4)$ chiral corrections.
The same approach provides a quite successful calculation of the
$K^+\to\pi^+\pi^0$ decay rate \cite{ref:GPR85},
in good agreement with the experimental value.
It overestimates this decay amplitude by less than 15\%
\cite{ref:JP91}.

There have been several QCD sum rule calculations based on
studies of three-point-function correlators
\cite{ref:CKKP86,ref:RY87,ref:BDG88,ref:PV91}.
After some initial disagreements, the final result seems
to be \cite{ref:BDG88} \footnote{
Larger values ($B_K \sim 0.75$-$1$) have been obtained in
Refs. \cite{ref:PV91}, where, following
Ref. \cite{ref:CKKP86}, only the non-factorizable piece is estimated.
This type of analysis has been criticized in Refs.
\cite{ref:RY87,ref:BDG88}.
}:
\be\label{eq:bk_3pf}
{B}_K(\mu^2) \, = \, 0.5\pm0.1\pm0.2
\qquad\qquad\mbox{\rm (QCD Sum Rules --3 Point Functions)}.
\ee
The theoretical uncertainty is rather large because the
perturbative gluonic corrections have not been included yet,
i.e. $\widehat{B}_K \sim B_K(\mu^2)$ with an arbitrary
renormalization-scale $\mu$.

Lattice calculations have given fluctuating results.
The first (statistically) accurate determinations gave results
compatible with vacuum saturation: $\widehat{B}_K =
0.88\pm0.20$ \cite{ref:GA88},
$1.03\pm0.07$ \cite{ref:BS90},
$0.77\pm0.07$ \cite{ref:KSGP90}.
It was realized later that finite-size effects were important
and the  extrapolation to the continuum limit could
decrease the final numerical results; depending on the assumed
extrapolation, $\widehat{B}_K = 0.66\pm0.06$ or $0.78\pm0.03$
(statistical errors only)
was obtained \cite{ref:GKS92}.
An improved investigation of the lattice spacing errors, has
recently given
the more precise value \cite{ref:SH94}:
\be\label{eq:bk_lattice}
\widehat{B}_K \, = \, 0.825\pm0.035 .
\qquad\qquad\mbox{\rm (Lattice)}.
\ee
This result has been obtained in the quenched approximation
and with degenerate quarks with mass $m_s/2$.
To assess the significance of these difficult calculations,
one should keep in mind the related $K^+\to\pi^+\pi^0$ decay
amplitude. Present lattice calculations of $\widehat{B}_K$
are still done in the $SU(3)$ limit; therefore, they also
provide a prediction for the $K^+\to\pi^+\pi^0$ decay rate.
At present, the lattice calculation overestimates the
$K^+$ decay amplitude by a factor of 2.

Table \ref{tab:bk} summarizes the different
calculations. Except for the lattice value, which is somewhat
bigger, all other results are in the range (\ref{eq:bk_general}).

\begin{table}
\begin{center}
\begin{tabular}{|c|c|c|}
\hline $\widehat{B}_K$ & Method & Reference
\\ \hline
$3/4$ & Leading $1/N_c$ & \protect\cite{ref:GTW81}  
\\
$0.37$ & Lowest-Order Chiral Perturbation Theory &
\protect\cite{ref:DGH82}
\\ \hline
$0.70\pm0.10$ & Next-to-Leading $1/N_c$ Estimate &
\protect\cite{ref:BBG88}
\\
$0.4\pm0.2$ & Next-to-Leading $1/N_c$ Estimate, $O(p^2)$ &
\protect\cite{ref:PR91}
\\
$0.42\pm 0.06$ & $\cO(p^4)$ CHPT + $1/N_c$ Estimate &
\protect\cite{ref:BR94}
\\ \hline
$0.39\pm0.10$ & QCD-Hadronic Duality &
\protect\cite{ref:PR85,ref:PDPPR91}
\\
$0.5\pm0.1\pm0.2$ & QCD Sum Rules (3-Point Functions) &
\protect\cite{ref:BDG88}
\\
$ 0.825\pm 0.035$ & Lattice (Quenched Approximation)
& \protect\cite{ref:SH94} \\ \hline
\end{tabular}
\end{center}
\caption{Values of $\widehat{B}_K$ obtained by various methods.}
\label{tab:bk}
\end{table}

\section{$B^0$-$\bar B^0$ matrix element}

The vacuum saturation approximation has been usually applied
to estimate the $\Delta B=2$ matrix element in Eq.~(\ref{eq:bb}),
i.e. $B_B(\mu^2)=1$ is  generally
assumed.
However, in contrast to the kaon system, this assumption
does not provide by itself an estimate of the hadronic parameter $\xi_B$,
because the $B^0$ decay constant has not been measured yet.

The theoretical determination of $f_B$ has a quite confusing history,
because many contradictory results have been published.
To a large extent, the discrepancies among the different analyses
stem from the different approximations assumed to be valid,
and/or the different input values used in the final numerics.

Owing to the large mass of the $b$ quark,
non-relativistic potential models were supossed to provide
a good estimate of $f_B$
\cite{ref:KR80,ref:CG90}. 
However, relativistic and short-distance
QCD corrections have been shown
\cite{ref:CG90,ref:MT89} 
to be very significant.
The meson decay constant is directly
proportional to the meson wavefunction at the origin;
thus, these calculations are very sensitive to the
assumed short-distance behaviour of the potential,
which explains the broad range of results obtained within
this approach  \cite{ref:KR80,ref:CG90,ref:MT89}.

In the infinite quark-mass limit,
the meson decay constant should scale as
\cite{ref:SV87} 
\be\label{eq:scaling}
f_P \,\toinf\, {\alpha_s(M_Q)^{1/\beta_1}\over\sqrt{M_Q}} ,
\ee
where $\beta_1 = (2 n_f -33)/6$ is the first coefficient of the
QCD $\beta$ function, and $n_f$ the number of unfrozen flavours.
Assuming the charm-quark mass to be heavy enough, one should then
expect $f_B/f_D\sim 0.6$. However, this is not supported
by modern QCD sum rules \cite{ref:NA87,ref:NA94,ref:NE92}
and lattice calculations \cite{ref:BE94}  
which rather
prefer $f_B\sim f_D$, or even $f_B\gsim f_D$.
The origin of this unexpected behaviour can be understood
analyzing the leading corrections to the asymptotic result
(\ref{eq:scaling}):
\be\label{eq:fb_exp}
f_P \, = \, f_P^{\mbox{\rms stat}} \, \left\{1 + {c_P\over M_Q}
+ O(1/M_Q^2)\right\} .
\ee
The so-called static limit of the decay constant,
$f_P^{\mbox{\rms stat}}$, can be calculated using heavy-quark
effective theory methods
\cite{ref:NE93,ref:NA87,ref:NA94,ref:NE92,ref:BE94,ref:BBBD92};
moreover, several estimates of the
$1/M_Q$ correction have been performed
\cite{ref:NE93,ref:NA94,ref:NE92,ref:BBBD92,ref:BA94}.
These studies have shown
that:
\begin{itemize}
\item The value of $f_B^{\mbox{\rms stat}}$ is quite large,
typically $f_B^{\mbox{\rms stat}} \sim 2 f_\pi$. If short-distance
QCD corrections are ignored,
a much smaller value is obtained
\cite{ref:NA87,ref:NA94,ref:RA91}; however,
owing to the Coulombic interaction between the light and heavy quarks,
there is a large\footnote{
The next-to-leading order renormalization group improvement of the
currents in the heavy-quark effective theory
\protect\cite{ref:NE93,ref:NE92,ref:BBBD92,ref:BG92}
shows that the strong coupling constant must be evaluated at
a characteristic low-energy hadronic scale of order 1 GeV,
rather than at the scale of
the heavy quark.
Thus, the relevant coupling $\alpha_s(\mu)$ is much larger than
the one used in previous analysis
\protect\cite{ref:ES92}, 
which results in a sizeable increase
of $f_B^{\mbox{\rms stat}}$.}
perturbative gluonic correction of order 100\% \cite{ref:BG92}.
\item The leading $1/M_Q$ correction is negative and sizeable,
$c_P \sim -1$ GeV. It amounts to a 20\%
decrease of the decay constant at the $b$-quark scale,
but it is of order 100\% at the charm-mass scale, which makes
a non-relativistic determination of $f_D$ meaningless.
\end{itemize}
These large corrections allow us to understand the
discrepancies among previous approximate calculations, but
at the same time point out the difficulty of making a reliable
determination of $f_B$ within a heavy-quark or non-relativistic
approach.
Anyhow, it has been argued \cite{ref:BBBD92}
that the large first-order gluonic
correction gives already a very good description of the classical
Coulomb interaction and, therefore,
there is no reason to expect additional large contributions
at higher-orders. Taking the pole $b$-quark mass to be
$m_b = 4.6 \, (4.8)$ GeV, values around
\cite{ref:NE93,ref:NE92,ref:BBBD92,ref:BA94}
$f_B \sim 1.4\, (1.1) f_\pi$ have been estimated, using
QCD sum rules in the heavy-quark effective theory.

The values of $f_B$ obtained with QCD sum rules have a sizeable
dependence on the input value of the heavy-quark mass. This effect
induces a large uncertainty and is to a large extent responsible
for the apparent discrepancy among different predictions
\cite{ref:NA87,ref:NA94,ref:DP87,ref:RE88,ref:MY84}.
Using the presently favoured range for the perturbative
pole mass \cite{ref:RE88,ref:NA89,ref:NA87b,ref:DP92b},
$m_b = (4.6\pm0.1)$ GeV,
one obtains \cite{ref:NA94}
\be\label{eq:fb_qcdsr}
f_B \, = \, (1.6 \pm 0.3)\, f_\pi
\qquad\qquad\qquad\mbox{\rm (QCD Sum Rules)},
\ee
where larger values of the decay constant correspond to lower
$b$-quark masses.
Taking a larger mass $m_b=4.8$ GeV, one gets instead
$f_B\sim f_\pi$, which shows
the strong sensitivity to the value of $m_b$.

Lattice simulations with propagating heavy quarks face the problem
of large systematic errors associated with the finite lattice
spacing. One can only consider quark masses such that
$M_Q a\lsim 1$ (for heavier masses, the associated
Compton wavelengths are smaller than the lattice spacing $a$),
which implies that only the region around the charm-quark mass
can be investigated with the presently available lattices.
An extrapolation to the $b$-quark scale is then unavoidable.
The currently quoted results \cite{ref:BE94} 
are in the range
\be\label{eq:fb_lattice}
f_B \, = \, (1.4\pm0.3)\, f_\pi
\qquad\qquad\qquad\mbox{\rm (Lattice)}.
\ee

Concerning the $B_B$ factor, one would naively expect that the
vacuum insertion approximation becomes more reliable with the increase
of the quark mass. There are two published studies of this quantity,
using QCD sum rules based on three-point-function correlators,
which favour indeed a value  of $B_B$ around one:
$B_B(\mu^2) = 0.95\pm0.10$ \cite{ref:RY88} 
However, perturbative $\alpha_s$ corrections have not been considered
and therefore the scale $\mu$ is arbitrary.
Moreover, what is actually computed is
$f_B^2\xi_B = f_B^4 B_B$; thus, the value of $f_B$
(or the sum rule used to fix the decay constant) is needed as input,
which increases the theoretical uncertainty of the calculation.

So far, only one direct estimate of the relevant
quantity $\xi_B$ has been published \cite{ref:PI88}.
This calculation is based on the two-point-function correlator
associated with the $\Delta B=2$ four-quark operator. The usual
QCD sum rule technology allows to fix the value of $\xi_B^2$
directly, without any prior knowledge of $f_B$.
Unfortunately, the result turns out to be very sensitive to the
input value of the bottom quark mass.
Taking the range $m_b = (4.6\pm0.1)$ GeV, one gets
\cite{ref:PI88}:
\be\label{eq:xib_sr}
\xi_B(\mu^2) \, = \, (1.7\pm0.4)\, f_\pi
\qquad\qquad\qquad\mbox{\rm (QCD Sum Rules)},
\ee
the larger $\xi_B$ values corresponding to the lower masses.
The comparison with Eq.~(\ref{eq:fb_qcdsr})
could give some support to the
vacuum saturation approximation. However this comparison
is not very meaningful because it is not specified at which
scale it refers;
again, perturbative gluonic contributions have not been included yet.
One can argue that the relevant scale in the QCD sum rule is
$\mu = 2 m_b$ \cite{ref:PI88}; the resummation
of leading logarithms gives then rise to the
renormalization-scale independent mixing parameter
[see Eq.~(\ref{eq:xi_b})]
\be\label{eq:xib_hat}
\hat\xi_B \, = \, (2.0\pm0.5) f_\pi .
\ee
Taking instead $\mu = m_b$ would not change the result within the
quoted error-bar.
Note that with $\mu\sim m_b,2m_b$, $\hat\xi_B$ is a 20\% larger
that $\xi_B(\mu^2)$.
To make a better estimate, a calculation of the
next-to-leading-logarithm corrections is needed.
Those corrections which are factorizable are already known from 
the analogous $f_B$ calculation, and produce \cite{ref:JP91} a
18\% increase with respect to $\hat\xi_B$ in (\ref{eq:xib_hat});
the non-factorizable contributions remain, however, unknown\footnote{
The non-factorizable contributions have been recently estimated
to be smaller than 15\% \protect\cite{ref:NP94}. Unfortunately,
this number refers to some special Fierz-symmetric
renormalization-scheme. It is not clear how to translate it
into the usual $\overline{MS}$ scheme, which is needed for consistency
with the rest of the theoretical analysis.}.

As in the $f_B$ case, lattice simulations have estimated the
parameter $B_P$ in the charm region.  Performing an extrapolation
to the physical $B$-meson mass,
the present lattice determination is \cite{ref:ELC92}
$\widehat{B}_B = 1.16\pm0.07$. Using the $f_B$ value obtained in the
same lattice, $f_B = (1.54\pm0.30)\, f_\pi$, the quoted result for the
renormalization-scale-independent parameter $\hat{\xi}_B$ is
\cite{ref:ELC92}:
\be\label{eq:xib_lattice}
\hat{\xi}_B \, = \, (1.7\pm 0.3)\, f_\pi
\qquad\qquad\qquad\mbox{\rm (Lattice)},
\ee
in agreement with the QCD sum rule value (\ref{eq:xib_hat}).

\section{Numerical analysis}

In order to perform a numerical analysis of the UT constraints, we
take
$\Lambda^{(5)}_{\overline{MS}}= 240 \pm 90$ MeV
\cite{ref:ALT92},
the charm quark pole mass $m_c= 1.47\pm 0.05$ GeV
\cite{ref:NA89},
$\lambda = 0.2205\pm0.0018$ and the inputs shown
in Table~\ref{tab:inputs}. In view of the uncertainties discussed
before, we have taken two different sets of parameters:
the first corresponds to our best estimate, while the second one
represents a somewhat more conservative choice.

\begin{table}[htb]
\begin{center}
\begin{tabular}{|c|c|c|}
\hline Parameter & Best estimate & Conservative choice
\\ \hline
$x_d$ &
$0.71 \pm 0.07$ &
$0.70 \pm 0.10$
\\
$m_t^{\mbox{\rms pole}}$ &
$174 \pm 16 \, \mbox{\rm GeV}$ &
$175 \pm 25 \, \mbox{\rm GeV}$
\\
$\tau(B^0_d)$ &
$1.49\pm 0.04 \, \mbox{\rm ps}$ &
$1.50\pm 0.10 \, \mbox{\rm ps}$
\\
$\tau(B^0_d)\, |V_{cb}|^2$ &
$(3.1\pm 0.7)\times 10^9\,\mbox{\rm GeV}^{-1}$ &
$(3.1\pm 1.0)\times 10^9\,\mbox{\rm GeV}^{-1}$
\\
$|V_{ub}|/|V_{cb}|$ & $0.08\pm0.02$ & $0.08\pm0.03$
\\
$\widehat{B}_K$ & $0.50\pm0.15$ & $0.55\pm0.25$
\\
$\hat\xi_B/ f_\pi$ & $2.0\pm 0.5$ & $1.9\pm 0.7$
\\ \hline
\end{tabular}
\end{center}
\caption{Input values for the UT analysis.}
\label{tab:inputs}
\end{table}

\begin{figure}
\vspace*{-4cm}
{\epsfysize=25.5cm\epsfxsize=17cm\epsfbox{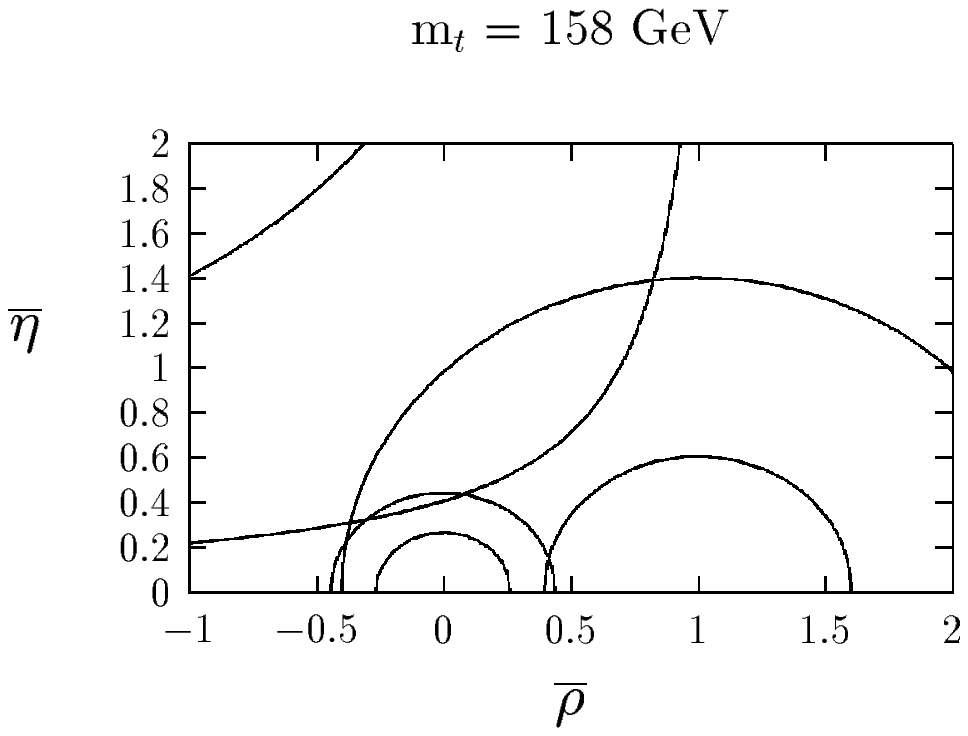}}
\vspace*{-19cm}
{\epsfysize=25.5cm\epsfxsize=17cm\epsfbox{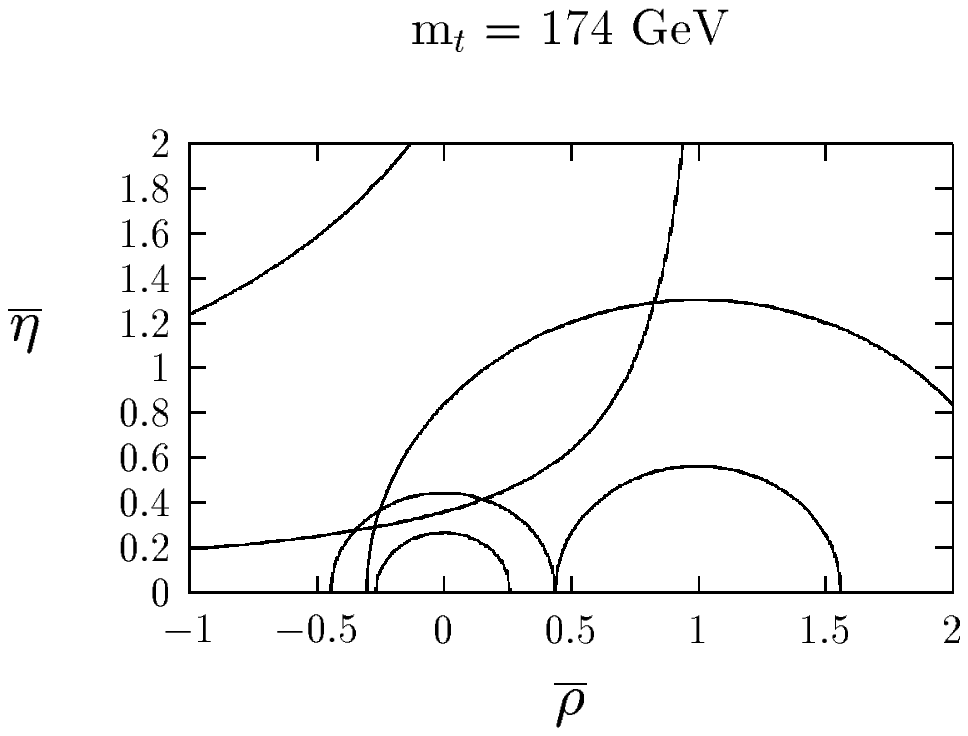}}
\vspace*{-19cm}
{\epsfysize=25.5cm\epsfxsize=17cm\epsfbox{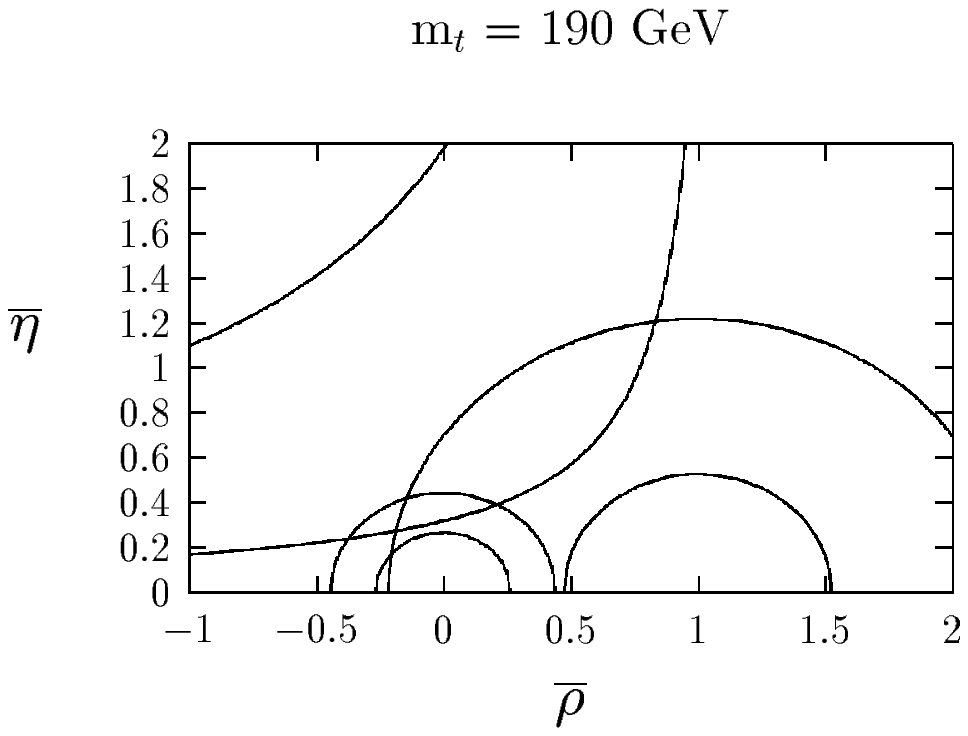}}
\vspace*{-15cm}
\caption{Constraints on the UT for the {\em best}
 estimate set of parameters in Table \protect{\ref{tab:inputs}}.}
\label{fig:set1}
\end{figure}

\begin{figure}
\vspace*{-5cm}
{\epsfysize=25.5cm\epsfxsize=17cm\epsfbox{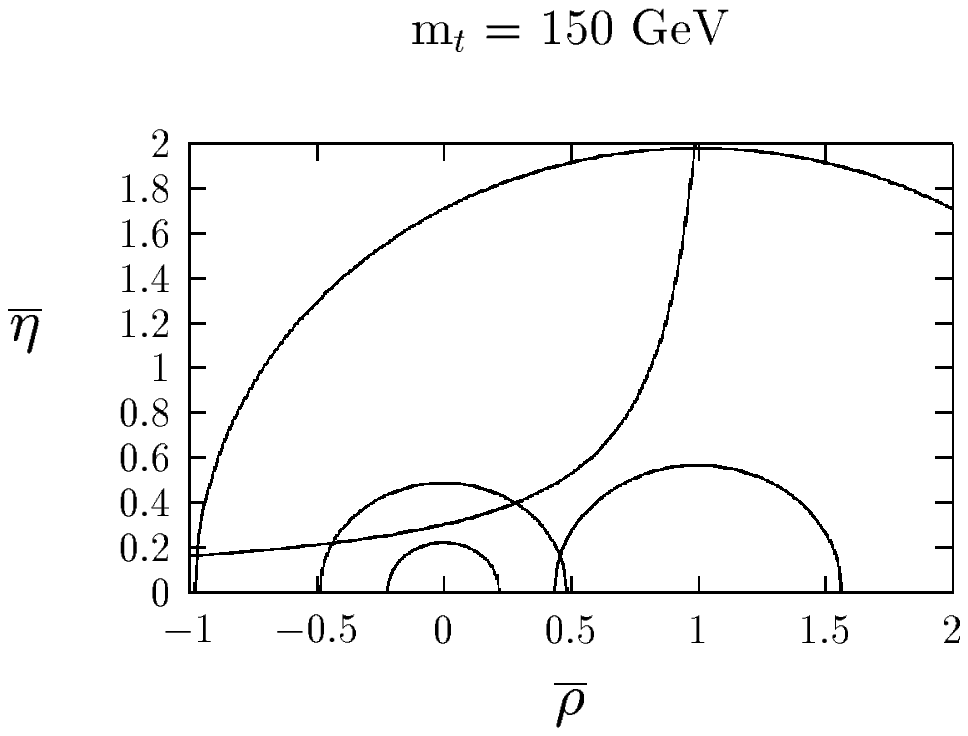}}
\vspace*{-19cm}
{\epsfysize=26cm\epsfxsize=17cm\epsfbox{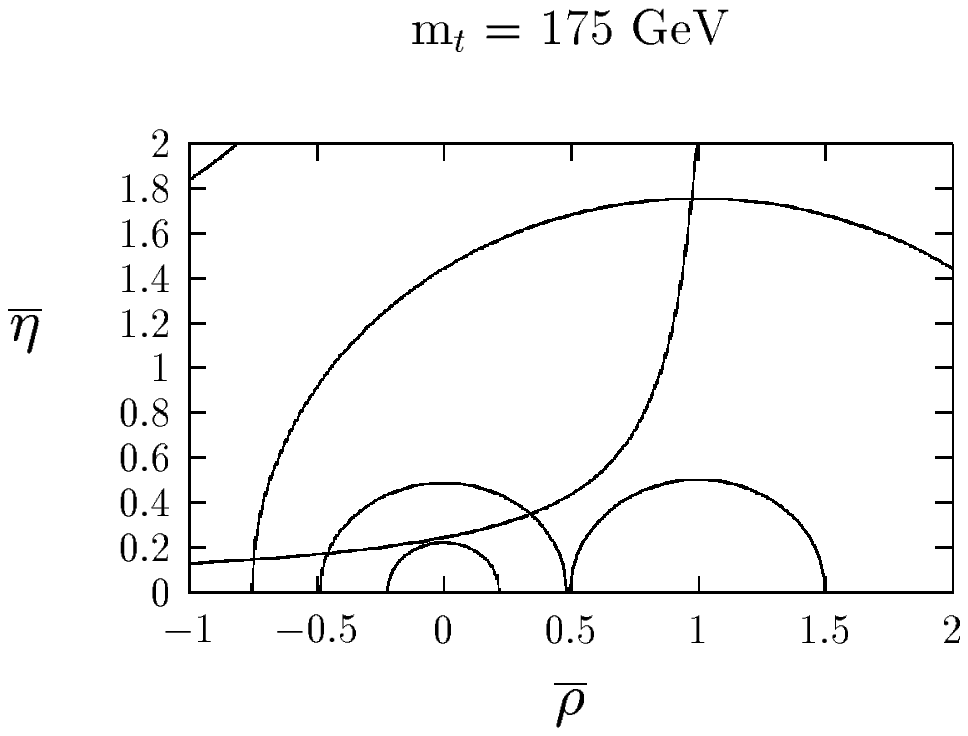}}
\vspace*{-19cm}
{\epsfysize=25.5cm\epsfxsize=17cm\epsfbox{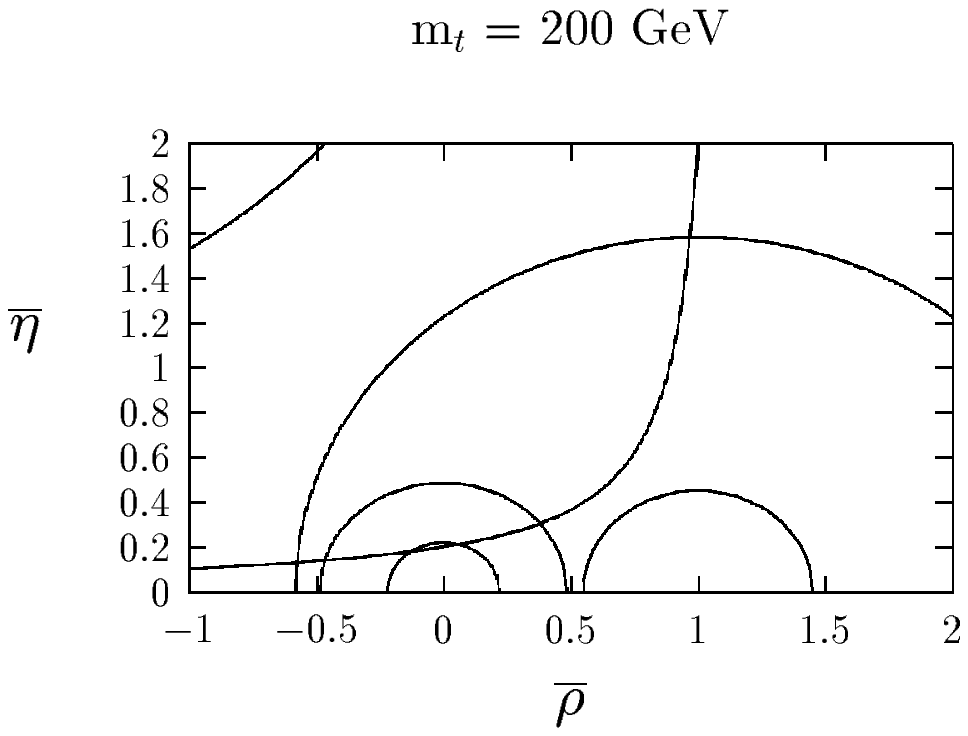}}
\vspace*{-15cm}
\caption{Constraints on the UT for the {\em conservative}
 choice  of parameters in Table \protect{\ref{tab:inputs}}.}
\label{fig:set2}
\end{figure}

Figures \ref{fig:set1} and \ref{fig:set2} show the resulting UT
constraints for the {\em best} and the
{\em conservative}, respectively, choices of input parameters.
The circles centered at $(0,0)$ show the
present determination of $R_b$, defined in Eq. (\ref{eq:rb}).
The measured $B^0$-$\bar B^0$ mixing parameter $x_d$,
constraints the value of
$R_t$, defined in Eq. (\ref{eq:rt}),
forcing the vertex $(\bar\rho,\bar\eta)$
to be in the region between the two
circles centered at $(1,0)$;
the bigger circle corresponds to the smaller allowed values
of $\hat\xi_B$ and $|V_{cb}|$.
The hyperbolae show the constraints from the
$K^0$-$\bar K^0$ CP-violating
parameter $| \varepsilon|$, which follow from Eq.
(\ref{eq:hyperbola}); smaller values of $\widehat{B}_K$
and  $|V_{cb}|$ correspond to
larger values of $\bar\eta$.
The final allowed range of values
for $(\bar \rho, \bar \eta)$ is given
by the area which is common to the regions in between the
hyperbolae, the circles centered at $(0,0)$ and the
circles centered at $(1,0)$.
%
%
In Tables \ref{tab:results1} and \ref{tab:results2}
we give the numerical results for $\sin (2 \alpha)$,
$\sin (2 \beta)$, $\sin (2 \gamma)$, $R_b$ 
and $R_t$ from our analysis of the UT, using the two sets of input
parameters introduced in Table \ref{tab:inputs}.

\begin{table}[htb]
\begin{center}
\begin{tabular}{|c|c|c|c|}
\hline Parameter & $m_t^{\mbox{\rms pole}} = 158$ GeV &
$m_t^{\mbox{\rms pole}} = 174$ GeV &
$m_t^{\mbox{\rms pole}} = 190$ GeV
\\ \hline
$\sin(2 \alpha)_{\mbox{\rms max}}$ &
$0.96$ &
$0.97$ &
$0.98$
\\
$\sin(2 \alpha)_{\mbox{\rms min}}$ &
$0.43$ &
$0.16$ &
$-0.10$
\\
$\sin(2 \beta)_{\mbox{\rms max}}$ &
$0.76$ &
$0.81$ &
$0.81$
\\
$\sin(2 \beta)_{\mbox{\rms min}}$ &
$0.43$ &
$0.44$ &
$0.43$
\\
$\sin(2 \gamma)_{\mbox{\rms max}}$ &
$0.41$ &
$0.68$ &
$0.83$
\\
$\sin(2 \gamma)_{\mbox{\rms min}}$ &
$-1.0$ &
$-1.0$ &
$-1.0$
\\
$R_b$ &
$0.42^{+0.03}_{-0.04}$ &
$0.39 \pm 0.07$ &
$0.38 \pm 0.07$
\\
$R_t$ &
$1.17 \pm 0.17$ &
$1.12 \pm 0.18$ &
$1.06 \pm 0.18$
\\ \hline
\end{tabular}
\end{center}
\caption{Numerical results for the {\em best} estimate set
of input parameters.}
\label{tab:results1}
\end{table}

\begin{table}[htb]
\begin{center}
\begin{tabular}{|c|c|c|c|}
\hline Parameter & $m_t^{\mbox{\rms pole}} = 150$ GeV &
$ m_t^{\mbox{\rms pole}} = 175$ GeV &
$m_t^{\mbox{\rms pole}} = 200$ GeV
\\ \hline
$\sin(2 \alpha)_{\mbox{\rms max}}$ &
$0.95$ &
$0.95$ &
$0.95$
\\
$\sin(2 \alpha)_{\mbox{\rms min}}$ &
$-0.25$ &
$-0.60$ &
$-0.80$
\\
$\sin(2 \beta)_{\mbox{\rms max}}$ &
$0.86$ &
$0.87$ &
$0.87$
\\
$\sin(2 \beta)_{\mbox{\rms min}}$ &
$0.29$ &
$0.25$ &
$0.18$
\\
$\sin(2 \gamma)_{\mbox{\rms max}}$ &
$1.0$ &
$1.0$ &
$1.0$
\\
$\sin(2 \gamma)_{\mbox{\rms min}}$ &
$-1.0$ &
$-1.0$ &
$-1.0$
\\
$R_b$ &
$0.40 \pm 0.10$ &
$0.36 \pm 0.14$ &
$0.36 \pm 0.14$
\\
$R_t$ &
$1.12 \pm 0.30$ &
$1.09 \pm 0.36$ &
$1.07 \pm 0.39$
\\ \hline
\end{tabular}
\end{center}
\caption{Numerical results for the {\em conservative} choice
of input parameters.}
\label{tab:results2}
\end{table}

For the {\em best} estimate set of input parameters we get
$\sin(2 \beta)>0.43$ while for the {\em conservative} choice
we have $\sin(2 \beta) > 0.29,0.25$ and 0.18,
for $m_t^{\mbox{\rms pole}} = 150, 175$ and 200 GeV, respectively.
In this last case, the constraints are mainly  imposed
by $R_b$ and the CP-violating parameter $\varepsilon$, as can be
seen from the figures. The difference between both estimates,
namely,
{\em best} versus {\em conservative},
gives a good idea of the present uncertainties.

Assuming the Standard Model mechanism of CP violation to be correct,
one can use the UT analysis to pin down the values of the relevant
input parameters. For instance, the measured low values of
$|V_{ub}|/|V_{cb}|$ imply that larger values of $\widehat{B}_K$
and/or $|V_{cb}|$
are preferred;
alternatively,  a low value of $\widehat{B}_K$
could indicate that the theoretical uncertainties in the analyses
of the decays $B\to D^*l\bar\nu_l$ and
$B\to X_q l \bar\nu_l$ have been underestimated.
Similarly, larger values of  $|V_{cb}|$  would favour smaller
values of $\hat\xi_B$ (for a given top mass).
However, this kind of analysis is not very satisfactory,
because it misses the main motivation for studying the unitarity relation
(\ref{eq:unitarity}): to test the Standard Model mechanism of CP violation.
If more precise experimental data shows some discrepancy between different
UT constraints, we would like to know if
a violation of CKM universality has been established (i.e. new physics),
or if the reason is just a wrong theoretical determination
of some (less interesting) non-perturbative parameter.
Clearly, a better theoretical understanding of long-distance effects
is needed.

\section*{Acknowledgements}

We would like to thank Hans Bijnens and Eduardo de Rafael for
a critical reading of the manuscript. This work has been supported
in part by CICYT (Spain) under Grant Nr. AEN-93-0234.

%

\end{document}